\documentclass[prb,superscriptaddress,twocolumn]{revtex4}
\usepackage{amsmath}
\usepackage{bm}
\usepackage{amssymb}
\usepackage{graphicx}

\def\bk{\bar k}
\def\tb{\tilde b}

\def\bsigma{\bar\sigma}
\begin{document}

\title{Multiscale Multiexciton Cyclic Dynamics in Light Harvesting Complex}
\author{Shmuel Gurvitz}
\affiliation{Department of Particle Physics and Astrophysics, Weizmann Institute,
76100, Rehovot, Israel}
\author{Gennady P.  Berman}
\affiliation{Theoretical Division, Los Alamos National Laboratory and the New Mexico Consortium, Los Alamos, NM, 87544, USA}
\author{Richard T. Sayre}
\affiliation{Bioscience Division, B-11, Los Alamos National Laboratory and the New Mexico Consortium, Los Alamos, NM, 87544, USA}

\begin{abstract}
	Usually the study of energy-transfer in the light harvesting complex is limited by a single-exciton motion along the antenna. Starting from the many-body Schr\"odinger equation, we derived  Lindblad-type Master equations describing the cyclic exciton-electron dynamics of the light harvesting complex, originated from charge reduction of a donor. These equations, resembling the Master equations for the electric current in mesoscopic systems, go beyond the single-exciton description by accounting for the multi-exciton states accumulated in the antenna, as well as the charge-separation, fluorescence and initial photo-absorption. Although these effects take place on very different time-scales, we demonstrate that their inclusion is necessary for a consistent description of the exciton dynamics. We applied our results to evaluate the energy (exciton) current and for the fluorescent current depending on the light-intensity.
\end{abstract}

\preprint{LA-UR-17-24343}
\maketitle

\section{Introduction}

\begin{figure}[h]
\includegraphics[width=8cm]{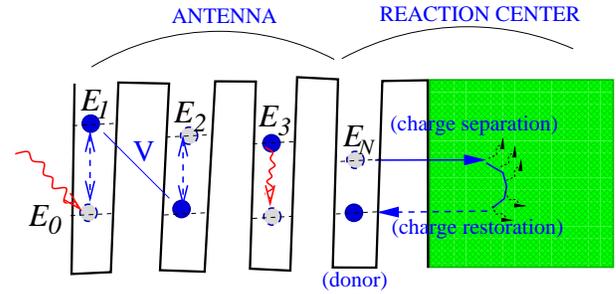}
\caption{An exciton, created on the left site of the antenna via photon absorption, moves along the  antenna, due to the dipole-dipole interaction, $V$, until it arrives at the site $N$ (RC ``donor''), where  primary charge separation takes place. After the charge separation and until the donor's  charge-neutrality is restored, no other excitons (if they occur in the LHC) can enter the RC donor. They would occupy other sites of the LHC, until they are lost by fluorescence or recombination processes.}
\label{fig1}
\end{figure}

The energy transfer in the light-harvesting antenna complex (LHC) takes place via exciton propagation among pigments bound to the LHC proteins \cite{book1}. The exciton is created by resonant photo-absorption on an antenna pigment, leading to electron excitation from the ground to the excited energy level, $\gamma +E_0 \to E_1$, Fig.~\ref{fig1}.  Due to the dipole-dipole interaction, $V$, the exciton then propagates between neighboring pigments to the reaction center (RC), while all, excited and non-excited sites of the antenna, remain  uncharged. Finally, the exciton arrives at the site $N$ (the ``donor'' of the RC), where the primary process of the charge (electron) separation occurs.  As a result the, donor becomes positively charged, and the electron participates in chemical reactions in the RC. Finally,  at some time $\tau$ the donor's charge is reduced by an electron ultimately derived from water oxidation, Fig.~\ref{fig1}, and the cycle is completed.

The dynamics of the exciton transfer along the antenna, including the primary charge separation is very rapid ($\sim$ ps). Otherwise, the exciton would be lost by fluorescence or by other (recombination) processes, taking place on the time-scale of $\sim$ ns. In comparison, the duration of entire cycle ($\tau$), completed with reduction of  the primary oxidized donor, is much longer ($\sim \mu s$). During the cycle no excitons occupy the RC donor. However, they can be accumulated by the antenna pigments, and finally being lost  by  fluorescence, as shown schematically in Fig.~\ref{fig1}.

Usually, the study of energy-transfer in the LHC is limited by a single-exciton migration along the antenna pigment bed. The initial photo-absorption, fluorescence and charge restoration of the RC donor are not included in the consideration, since they occur at significantly different time-scales. Nevertheless, without a consistent accounting of all these multi-scale processes one cannot fully understand and describe the exciton dynamics in the LHC. For instance, for efficient performance of the LHC, no more than one photon can be absorbed during the RC cycle, $\tau$, corresponding to the optimal absorption rate, $\sim 1/\tau$.  However, if the intensity of the sunlight increases, more than one exciton can accumulate inside the antenna. These ``trapped"  excitons can damage the photosynthetic apparatus through de-excitation pathways leading to the generation of oxygen singlets. In competition with this destructive pathway is de-excitation of excess excited states by non-photochemical quenching (NPQ) mechanisms \cite{book2}. It is clear that, in order to understand the NPQ mechanism, energy transfer from the LHC to the RC cannot be treated in terms of a single-exciton transport, during one cycle.

\begin{figure}[h]
\includegraphics[width=7cm]{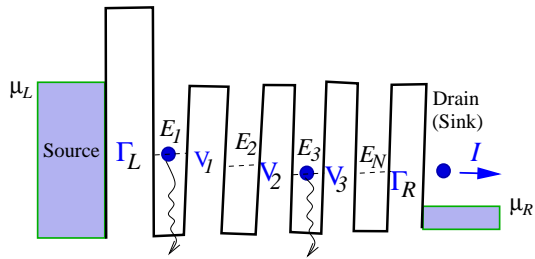}
\caption{Electron transport through coupled quantum dots under external bias, $\mu_L-\mu_R$, where $\mu_{L,R}$ are chemical potentials. Each electron in the system can leak to another reservoir, similar to the loss of excitons due to fluorescence shown in Fig.~\ref{fig1}.}
\label{fig11}
\end{figure}

At first sight, exciton transport along the LHC appears similar to spinless electron transport in a mesoscopic system. Indeed, no more than one exciton can reside on the same site, if only one excitation is allowed for each site (hard exciton model) \cite{muc}. As a result, the exciton propagation along the antenna would be similar to electron tunneling through coupled-dot system, as shown schematically in Fig.~\ref{fig11}. Here, the left and the right leads (source and drain), with tunneling rates, $\Gamma_{L,R}$, play the role of the initial photon flux and the RC, Fig.~\ref{fig1},

Note, that one can assume that $\Gamma_L\ll \Gamma_R$, so that no more than one electron enters the system (similar to the LHC exciton transport in normal regime). The electron would make coherent oscillation between the dots, unless it arrives to the drain (sink). Then,  it cannot return back, since the number of available states in the sink is very large. Also, the probability of its return to the source lead would be negligible with respect to the incoming flux of electrons. This also could explain the irreversibility of the exciton current \cite{farrow}. Therefore, such an analogy between the exciton and electron transport can be helpful. One can also introduce a coupling of each dot with another (leakage) reservoir which would be analogous to a loss of excitons due to fluorescence in the LHC.

The treatment of electron current through the coupled dots, Fig.~\ref{fig11}, can be greatly simplified in the case of a large bias voltage $(\mu_L-\mu_R\gg\Gamma, V)$. Then, the many-body Schr\"odinger equation $i|\dot\Psi(t)\rangle =H|\Psi(t)\rangle$, where $H$ is the total Hamiltonian, can be reduced to the Lindblad-type particle-number-resolved Master equations, beyond the commonly used weak coupling approximation,
see Ref.~[\onlinecite{gp}] and recent review paper~[\onlinecite{gur}], with detailed derivations. Using this method, one can evaluate in a simple way  the electron current through the system, together with the leakage electron current.

It is desirable to realize this analogy and derive similar Master equations for the exciton transport in the LHC, Fig.~\ref{fig1}. However, in this case we have to include in the Hamiltonian, $H$, terms describing the primary charge separation, which takes place at the RC donor site, and the restoration of the primary donor's neutrality.

The primary charge separation can be considered as an electron transition from the excited donor state, $E_N$, to the acceptor (charge) states, $E_{C_i}$. The  process follows by the charge stabilization \cite{sayre}, which implies that the electron is not returning to the initial donor state, $E_N$. Therefore, the acceptor cannot be represented by a single state. Otherwise, the electron would oscillate between the donor and the acceptor. For this reason we have to consider the acceptor to be a band of dense levels, $E_{C_i}$ (sink), centered around $E_N$. Then, the primary charge separation is represented by tunneling of the electron from the donor to the sink with a rate, $\Gamma\sim$1/ps.

\begin{figure}[h]
\includegraphics[width=8cm]{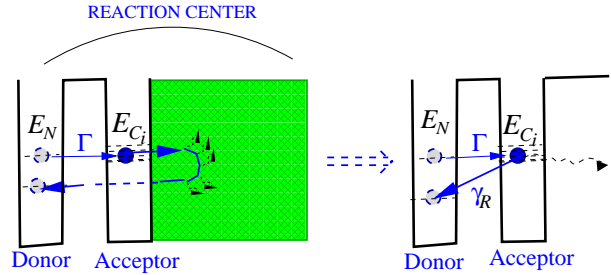}
\caption{Schematic representation of the electron cycle, from a primary charge separation from the donor, $E_N$, to the acceptor, $E_{C_i}$. The cycle is completed with restoration of the donor's  neutrality. The latter is modeled as a relaxation to the donor's ground state, by emission of energy (fictitious boson) in the RC.}
\label{fig2}
\end{figure}

In order to describe the restoration of the primary donor's neutrality in a phenomenological way, it is not necessary to know all details of the slow chemical reactions in the RC, initiated by the electrons. What is relevant, is only the period of the cycle ($\tau$). Thus, we can effectively represent the charge restoration as a slow relaxation of an electron from the acceptor to the donor's ground state, accompanied by the emission of the energy ($E_N-E_0$) in the RC, Fig.~\ref{fig2}. Although the  cycle is completed by an electron coming from a different place (such as water splitting in the photosystem II, etc.), its origin is not important for the exciton dynamics in the LHC.

The relaxation process cannot be described by considering the sink ($E_{C_i}$) as only an effective Markovian reservoir, absorbing an electron after the primary charge separation. Indeed, the electron cannot return from the reservoir to the (localized) ground state with large energy transfer.
Rather, the relaxation can be modeled by emission of a fictitious particle (boson) carrying this energy. For this reason, we introduce in the total Hamiltonian a very weak coupling of each electron state in the sink with a bath of fictitious bosons. If the bath is initially empty, the electron would exponentially decay to the donor's ground state with a relaxation  decay rate, $\gamma_R=1/\tau$, by emitting
 fictitious bosons with the energy, $E_{N}-E_0$, Fig.~\ref{fig2}.

Inclusion of a fictitious bosons together with a quantized electro-magnetic field, describing the initial light (photon) source and fluorescence, into the total Hamiltonian, would allow us to derive closed Master equations for exciton dynamics in a complete quantum mechanical way. As a result, we would be able to evaluate the energy (exciton) current as a function of the incoming sunlight intensity and the probability of the multi-exciton states inside the antenna. Dissipation of these multi-excitons by fluorescence is studied as well, by evaluation of the fluorescent current. This approach can be considered as a framework for the treatment of the energy transfer through any network in the antenna complex and also in the presence of noise. The latter, however,  will be studied in a separate work.

The paper is organized as follows. In Sec. IIA, we present a derivation of the Master equations for photo-absorption by a single electron site, using our wave-function approach. In this way, we reproduce old results, existing in the literature, but obtained in a different way.  Primary charge separation and restoration on the donor site is investigated in Sec. IIB. Section IIC presents the Master equations, describing a complete cycle for one site antenna. Simple analytical expressions for the energy current and for the probability of different exciton states are obtained. Sec. III deals with the general case of an $N$-site antenna. The Hamiltonian and the general particle-number-resolved Master equations are presented in Sec. III A. The examples of two and three-site antenna are discussed in detail in Sec. III B,C. A particular attention is paid to the fluorescent current and to its dependence on the sunlight intensity. Last section is the Summary.

\section{Derivation of Master equations for one-site antenna}

\subsection{Rate equations for photo-absorption}

Let us derive Master equations for photo-absorption on the first (peripheral) site of antenna, Fig.~\ref{fig1}, separated from the rest of antenna. The corresponding Hamiltonian, $H_1$, can be written as,
\begin{align}
H_1=E_0\hat a_{0}^\dagger \hat a_{0}^{}&+E_1\hat a_1^\dagger \hat a_1^{}+\sum_k\omega_k\hat C_k^\dagger \hat C_k^{}\nonumber\\
&~~~~~~+\sum_k\big(g_k\hat B_1^\dagger\hat C_k+H.c.\big),
\label{apb1}
\end{align}
where $\hat a_1^\dagger(\hat a_1^{})$ is an electron creation (annihilation) operator for the excited state, $E_1$, and $\hat a_{0}^\dagger (\hat a_{0}^{})$ is the same for the ground state, $E_0$, (in following we take $E_0=0$), while $\hat C_k^\dagger (\hat C_k^{})$ is a photon creation (annihilation) operator. $\hat B_1^{\dagger}=\hat a_1^{\dagger}\hat a_{0}^{}$ denotes an exciton creation operator. The last term in (\ref{apb1}) describes the electron-photon interaction in the rotating-wave approximation.

The time-dependent wave-function of the system is obtained from the Scr\"odinger equation, $i\partial_t|\Psi (t)\rangle =H_1|\Psi (t)\rangle$, where the initial state of the photon bath is a mixture of pure states,
\begin{align}
|\bar 0\rangle =\prod_{\bk}{(\hat C_{\bk}^\dagger)^{n_{\bk}^{}}\over\sqrt{n_{\bk}^{}!}}\hat a_0^\dagger|0\rangle\equiv \prod_{\bk}\hat a_0^\dagger|n_{\bk}\rangle,
\label{apb2}
\end{align}
and the index $\bk$ denotes the initially occupied states of the photon bath. Note that electron relaxation from higher to lower level, generated by the rotated-wave Hamiltonian~(\ref{apb1}), is accompanied by one-photon emission, while the reverse process is accompanied by one-photon absorption. As a result, the many-particle wave function, describing the evolution of the entire system can always be represented as,
\begin{align}
&|\Psi (t)\rangle = \Big [ b_0(t) + \sum_{\bk} b_{1\bk}(t)
{\hat B_{1}^{\dagger}\hat C_{\bk}^{}\over\sqrt{n_{\bk}^{}}}
           + \sum_{\bk,k} b_{\bk k}(t){\hat C_{k}^{\dagger}\hat C_{\bk}^{}\over\sqrt{n_{k}^{}+1}\sqrt{n_{\bk}^{}}}\nonumber\\
&+ \sum_{\bk <\bk',k} b_{1\bk\bk' k}(t)
{  B_{1}^{\dagger}\hat C_{k}^{\dagger}\hat C_{\bk}^{}\hat C_{\bk'}^{}\over\sqrt{n_{k}^{}+1}\sqrt{n_{\bk}^{}}\sqrt{n_{\bk'}^{}}}
           + \ldots \Big ] |\bar 0\rangle,
\label{wf}
\end{align}
where, $b_0(t)$, $b_{1\bk}(t)$, $b_{\bk,k}(t)$, and $b_{1\bk\bk' k}(t)$, are the probability amplitudes to find the system in the corresponding state, defined by the associated creation and annihilation operators. Here $\bk, k$ denote the states of absorbed and emitted photons. The initial conditions corresponds to $b_0(0)=1$, and all the other $b(0)$'s being zeros.

Substituting Eq.~(\ref{wf}) into the time-dependent Schr\"odinger equation, $i\partial_t|\Psi (t)\rangle =H_1 |\Psi (t)\rangle$, we find an infinite set of coupled equations for amplitudes, $b(t)$,
\begin{subequations}
\label{apb4}
\begin{align}
&i\dot{b}_{0}^{}(t) =\sum_{\bk}\sqrt{n_{\bk}^{}}g_{\bk}^{}
b_{1\bk}^{}(t)  \label{apb4a}, \\
&i\dot{b}_{1\bk}(t) =(E_{1}^{}-\omega_{\bk}^{})b_{1\bk}^{}(t)+\sqrt{n_{\bk}}g_{\bk}b_{0}(t)\nonumber\\
&~~~~~~~~~~~~~~~~~~~~~~~~~~~~~~~+\sum_k\sqrt{n_k^{}+1}g_k^{}b_{\bk k}^{}(t),  \label{apb4b}\\
&i\dot{b}_{\bk k}(t) =(\omega_{k}-\omega_{\bk})b_{\bk k}(t)+\sqrt{n_{k}+1}g_{k}^{}b_{1\bk}(t)\nonumber\\
&~~~~~~~~~~~~~~~~~~~~~~~~~~~~~~+\sum_{\bk'}\sqrt{n_{\bk'}}g_{\bk'}^{}b_{1\bk\bk' k}^{}(t),  \label{apb4c}\\
&i\dot{b}_{1\bk\bk' k}(t) =(E_{1}^{}+\omega_{k}^{}-\omega_{\bk}^{}-\omega_{\bk'}^{})b_{1\bk\bk' k}^{}(t)\nonumber\\
&~~~~~~~~~~~~~
+\sqrt{n_{\bk'}}g_{\bk'}b_{\bk k}(t)+\sqrt{n_{\bk}}g_{\bk}b_{\bk' k}(t)\nonumber\\
&~~~~~~~~~~~~~
+\sum_{k'}\sqrt{n_{k'}^{}+1}g_{k'}^{}b_{\bk\bk' kk'}^{}(t)  \label{apb4d},\\
&~~~~~~~~~~~~~~~~~~~~~~~~~~~~~\cdots\nonumber
\end{align}
\end{subequations}
For simplicity, we assumed that the photon-electron coupling, $g_k\equiv g(\omega_k)$, is real.

Using the Laplace transform,
\begin{align}
\tilde b(E)=\int\limits_0^\infty b(t)e^{iEt}dt,
\label{laplace}
\end{align}
one can rewrite Eqs.~(\ref{apb4}) as,
\begin{subequations}
\label{apbb4}
\begin{align}
&E{\tb}_{0}^{}(E) -\sum_{\bk}\sqrt{n_{\bk}^{}}g_{\bk}^{}
\tb_{1\bk}^{}(E)=i  \label{apbb4a}, \\
&(E+\omega_{\bk}^{}-E_{1}^{})\tb_{1\bk}^{}(E)-\sqrt{n_{\bk}}g_{\bk}\tb_{0}(E)\nonumber\\
&~~~~~~~~~~~~~~~~~~~~~~~~~~-\sum_k\sqrt{n_k^{}+1}g_k^{}\tb_{\bk k}^{}(E)=0,  \label{apbb4b}\\
&(E+\omega_{\bk}-\omega_{k})b_{\bk k}(E)-\sqrt{n_{k}+1}g_{k}^{}\tb_{1\bk}(t)\nonumber\\
&~~~~~~~~~~~~~~~~~~~~~~~~~-\sum_{\bk'}\sqrt{n_{\bk'}}g_{\bk'}^{}\tb_{1\bk\bk' k}^{}(E)=0,  \label{apbb4c}\\
&(E +\omega_{\bk}^{}+\omega_{\bk'}^{}-E_{1}^{}-\omega_{k}^{})\tb_{1\bk\bk' k}^{}(E)\nonumber\\
&~~~~~~~~~~
-\sqrt{n_{\bk'}}g_{\bk'}\tb_{\bk k}(E)-\sqrt{n_{\bk}}g_{\bk}\tb_{\bk' k}(E)\nonumber\\
&~~~~~~~~~~~~~~~~~
-\sum_{k'}\sqrt{n_{k'}^{}+1}g_{k'}^{}\tb_{\bk\bk' kk'}^{}(E)=0,  \label{apbb4d}\\
&~~~~~~~~~~~~~~~~~~~~~~~~~~~~~\cdots\nonumber
\end{align}
\end{subequations}
where the r.h.s. reflects the initial condition.

To simplify these equations, we use the same technique as developed in Refs.~[\onlinecite{gp,gur}] for the electron transport through quantum dots. It consists in a replacement of all sums by integrals, $\sum_k\to\int\rho(\omega_k) d\omega_k$, where $\rho$ is the density of photon states. Then, substituting the amplitude, $b(E)$, in each of the sums by an expression obtained from resolving the subsequent equation, we can perform an integration over the photon energy ($\omega_k$) analytically in the case of a wide band-width limit (Markovian reservoir). This implies that the spectral function, $|g(\omega_k)|^2\rho(\omega_k)$, is weakly dependent on $\omega_k$ and $E_1\gg g^2\rho$. However, no weak coupling limit is needed. The procedure is almost identical to that described in details in Sec 2 of Ref.~[\onlinecite{gur}], providing that electrons are replaced by photons. The latter results in the additional factors ($\sqrt{n_k^{}}$ and $\sqrt{n_k^{}+1}$) in front of the coupling constant, $g_k^{}$.

Proceeding with the  same algebra, as in Ref.~[\onlinecite{gur}], and then performing the inverse Laplace transform,
\begin{align}
b_{0,1}(t)=\int\limits_{-\infty}^\infty \tilde b_{0,1}(E)e^{-iEt}{dE\over 2\pi},
\label{invlap}
\end{align}
we transform Eqs.~(\ref{apbb4}), to the following equations,
\begin{subequations}
\label{apbb5}
\begin{align}
&i\dot{b}_{0}^{}(t) = i{\Gamma_{in}\over2}{b}_{0}^{}(t),\label{apbb5a} \\
&i\dot{b}_{1\bk}(t) =\Big(E_{1}^{}-\omega_{\bk}^{}+i{\Gamma_{out}\over2}\Big)b_{1\bk}^{}(t)
+\sqrt{n_{\bk}}g_{\bk}b_{0}(t), \label{apbb5b}\\
&i\dot{b}_{\bk k}(t) =\Big(\omega_{k}-\omega_{\bk}+i{\Gamma_{in}\over2}\Big)b_{\bk k}(t)+\sqrt{n_{k}+1}g_{k}^{}b_{1\bk}(t), \label{apbb5c}\\
&i\dot{b}_{1\bk\bk' k}(t) =\Big(E_{1}^{}+\omega_{k}^{}-\omega_{\bk}^{}-\omega_{\bk'}^{}
+i{\Gamma_{out}\over2}\Big)b_{1\bk\bk' k}^{}(t)\nonumber\\
&~~~~~~~~~~~~~~~~~~~~~~+\sqrt{n_{\bk'}}g_{\bk'}b_{\bk k}(t)+\sqrt{n_{\bk}}g_{\bk}b_{\bk' k}(t),  \label{apbb5d}\\
&~~~~~~~~~~~~~~~~~~~~~~~~~~~~~\cdots\nonumber
\end{align}
\end{subequations}
where
\begin{align}
\Gamma_{in}=\bar n\gamma~~{\rm and}~~\Gamma_{out}=(\bar n+1)\gamma,
\label{gammas}
\end{align}
with $\gamma=2\pi g^2(E_1)\rho (E_1)$, and $\bar n=n(E_1)$. Thus, $\Gamma_{in}$ is a rate of a photo-absorption, leading to electron transition from the ground to the excited state (exciton creation), and $\Gamma_{out}$ is a rate of a photo-emission in a reverse process (exciton annihilation).

Equations~(\ref{apbb5}) can be transformed to equations for the exciton density matrix, $\sigma(t)$, defined as,
\begin{align}
\sigma_{00}^{}(t) &= |b_{0}(t)|^2 + \sum_{\bk,k} |b_{\bk k}(t)|^2
            + \sum_{\bk<\bk'\atop k<k'} |b_{\bk\bk' kk'}(t)|^2 + \cdots\nonumber\\
             &\equiv \sigma_{00}^{(0)}(t)+\sigma_{00}^{(1)}(t)+\sigma_{00}^{(2)}(t)+\cdots,
             \nonumber\\[5pt]
\sigma_{11}^{}(t) &= \sum_{\bk} |b_{1\bk}(t)|^2 +
             \sum_{\bk<\bk',k} |b_{1\bk\bk' k}(t)|^2\nonumber\\
             &~~~~~~~~~~~~~~~~~~
             + \sum_{\bk<\bk'<\bk''\atop k<k'} |b_{1\bk\bk'\bk'' kk'}(t)|^2 + \cdots\nonumber\\
             &\equiv \sigma_{11}^{(0)}(t)+\sigma_{11}^{(1)}(t)+\sigma_{11}^{(2)}(t)+\cdots,
\label{sigmas}
\end{align}
where, $\sigma_{00}^{(p)}(t)$ and $\sigma_{11}^{(p)}(t)$, are probabilities  of finding the electron in the ground state and in the excited state (exciton), with $p$ photons emitted by time $t$. Respectively, $\sigma_{00}(t)$ and $\sigma_{11}(t)$ are total probabilities ($\sigma_{00}(t)+\sigma_{11}(t)=1$).

The reduction of Eqs.~(\ref{apbb5}) to rate equations for the density matrix, $\sigma_{jj}^{(p)}(t)$, defined by Eqs.(\ref{sigmas}), can be done straightforwardly (see Refs.[\onlinecite{gp,gur}]) by multiplying each of the equations (\ref{apbb5}) by a corresponding complex conjugated amplitude, and replacing sums in (\ref{sigmas}) by integrals. As a result, we arrived to the following rate equations,
\begin{subequations}
\label{aandb}
\begin{align}
&\dot{\sigma}^{(0)}_{00}(t) = - \Gamma_{in} \sigma^{(0)}_{00}(t)\;,
\label{anought}\\
&\dot{\sigma}^{(0)}_{11}(t) = \Gamma_{in} \sigma^{(0)}_{00}(t)
                               - \Gamma_{out} \sigma_{11}^{(0)}(t)\;,
\label{bnought}\\
&\dot{\sigma}^{(1)}_{00}(t) = - \Gamma_{in} \sigma^{(1)}_{00}(t)
                               + \Gamma_{out} \sigma_{11}^{(0)}(t)\;,
\label{aone}\\
&\dot{\sigma}^{(1)}_{11}(t) = \Gamma_{in} \sigma^{(1)}_{00}(t)
                               - \Gamma_{out} \sigma_{11}^{(1)}(t)\;,
\label{bone}\\
&~~~~~~~~~~~~~~~~~~~\cdots
\nonumber
\end{align}
\end{subequations}
which represent the particle-number-resolved Master equations of a form,
\begin{subequations}
\label{phot1}
\begin{align}
&\dot{\sigma}^{(p)}_{00}(t) = - \Gamma_{in} \sigma^{(p)}_{00}(t)+\Gamma_{out} \sigma^{(p-1)}_{00}(t)\;,
\label{phot1a}\\
&\dot{\sigma}^{(p)}_{11}(t) =  \Gamma_{in} \sigma^{(p)}_{00}(t)-\Gamma_{out} \sigma^{(p)}_{11}(t).
\label{phot1b}
\end{align}
\end{subequations}

These equations are identical to those describing the electron transport from the source to the drain through a single quantum dot, with $\Gamma_{in}$ and $\Gamma_{out}$ corresponding to the incoming and outgoing electron rates \cite{gp,gur}. Summing up these equations over $p$, one easily finds the rate equations for total probabilities,
\begin{subequations}

\label{apb8}
\begin{align}
& \dot{\sigma}_{00}(t)=-\Gamma_{in}\,\sigma_{00}(t)
+\Gamma_{out}\,\sigma_{11}(t) \label{apb8a}, \\
& \dot{\sigma}_{11}(t)=-\Gamma_{out}\,\sigma_{11}(t)
+\Gamma_{in}\,\sigma_{00}(t), \label{apb8b}
\end{align}
\end{subequations}
which can be rewritten as one equation,
\begin{align}
\dot{\sigma}_{00}(t)=-(2\bar n+1)\gamma\sigma_{00}(t)+(\bar n+1)\gamma,
\label{apb9a}
\end{align}
by taking into account that,  $\sigma_{11}(t)=1-\sigma_{00}(t)$.

In the steady state limit, $\dot\sigma_{00}(t\to\infty)\to 0$, so the ground state occupation, $\bar\sigma_{00}=\sigma_{00}(t\to\infty)$ is,
\begin{align}
\bar\sigma_{00}={\bar n+1\over 2\bar n+1}.
\label{st}
\end{align}
Therefore $\bar\sigma_{00}=1$ for $\bar n=0$, corresponding to fully occupied ground state. However, with increase of $\bar n$ (sunlight intensity), the occupation of the ground state decreases. In the limit $\bar n\to\infty$, corresponding to high sunlight intensity, $\bar\sigma_{00}\to 1/2$.

If the photon bath is in the thermal equilibrium state, then
$\bar n=1/(e^{E_1/T}-1)$. As a result, the occupation of the ground state is,
\begin{align}
\bar\sigma_{00}={1\over  1+e^{-E_1/T}},
\end{align}
which is a quite known result.  \cite{weiss,kor1}.

\subsection{Primary charge separation and restoration by emission of fictitious bosons.}

Consider the site $N$ (``RC donor'') of antenna, where the primary charge separation takes place, Figs.~\ref{fig1}, \ref{fig2}. As a result, an electron, occupying the excited level, $E_N$, of the RC donor is transferred to a neighboring site (``acceptor''), leaving the donor positively charged. This process is very fast ($\sim$ ps) in a comparison with the time-scales of the subsequent chemical reactions in the RC ($\sim \mu s$), generated by the electron. The cycle is complete, when the positively charged site, $N$, is neutralized (reduced) by an electron.

In contrast to the photo-absorption of the previous section, we treat these processes  phenomenologically. The first one, corresponding to the primary charge separation and stabilization, is considered as an electron tunneling from the donor to the acceptor site. However, the latter cannot be represented as a site containing isolated levels, since then the electron cannot be trapped at the acceptor: it will periodically returning to the donor.  In order to achieve the charge stabilization on the acceptor's site, we represent it as a sink, with a dense levels, $E_{C_i}$, centered around $E_N$. Then, the electron tunneling to the acceptor is trapped there. The corresponding tunneling rate, $\Gamma$, is very large ($\sim$ 1/ps).

The second process is modeled by a relaxation of the electron from the acceptor's band, $E_{C_i}$, to the donor's ground state with a very low rate, $\gamma_R$. It takes place due to emission of  fictitious bosons, representing the energy transfer to the RC. In order to describe these processes quantum-mechanically, we introduce an effective Hamiltonian, $H_N$, for the donor, $N$, separated from the rest of antenna, Fig.~\ref{fig2}. It can be written as,
\begin{align}
&H_N=E_N^{}\hat a_N^\dagger \hat a_N^{}+\sum_iE_{C_i}^{}\hat a_{C_i}^\dagger\hat a_{C_i}^{}+\sum_p\bar\omega_p^{}\hat F_p^\dagger \hat F_p^{}\nonumber\\
&+\sum_i\Big(\tilde V_i^{}\hat a_{C_i}^\dagger \hat a_N^{}
+\sum_pf_{ip}^{}\hat a_{0}^{\dagger}\hat a_{C_i}\hat F_p^{\dagger}+H.c.\Big).
\label{ham1}
\end{align}
Here, $\hat a_N^\dagger$ and $\hat a_{C_i}^\dagger$, denote electron creation operators at the site, $N$, and at a sub-level ($i$) of the acceptor,  $C$. Respectively, $\hat a_{0}^{\dagger}\equiv \hat a_{0N}^{\dagger}$, is an electron creation operator at the ground state of the donor (we chose $E_0\equiv E_{0N}=0$). The operator, $\hat F_p^{\dagger}$, describes a creation of  fictitious bosons, bearing the energy transferred to the RC. $\tilde V_i^{}$ is a tunneling coupling between the donor, $E_N$, with the sub-level, $E_{C_i}$, of the acceptor, and $f_{ip}^{}$ is a coupling of an electron on the acceptor with fictitious bosons.

The wave function, describing a whole system (electron and fictitious bosons) can be written as,
\begin{align}
|\Psi (t)\rangle &=\Big[b_N^{}(t)\hat a_N^\dagger a_{0}^{}+\sum_ib_{C_i}^{}(t)\hat a_{C_i}^\dagger\hat a_{0}^{}+\sum_p b_{0p}^{}(t)\hat F_p^\dagger\Big]|\bar 0\rangle,
\label{wf0}
\end{align}
where $|\bar 0\rangle\equiv \hat a_{0}^{\dagger}|0\rangle$ is the initial (``vacuum'') state of the system, corresponding to empty boson bath and the electron, occupying the donor's ground state.

Substituting Eq.~(\ref{wf0}) into the Schr\"odinger equation, $i\partial_t|\Psi (t)\rangle =H_N |\Psi (t)\rangle$, we find the following  system of coupled equations for the amplitudes $b(t)$,
\begin{subequations}
\label{eq0}
\begin{align}
&i\dot{b}_{N}^{}(t)=E_N^{}{b}_{N}^{}(t)+\sum_i \tilde V_i^{} b_{C_i}^{}(t),\label{eq0a}\\
&i\dot{b}_{C_i}^{}(t) =E_{C_i}^{}b_{C_i}^{}(t)+\tilde V_i^{} b_{N}^{}(t)+\sum_pf_{ip}^{} b_{0p}^{}(t)\label{eq0b},\\
&i\dot b_{0p}^{}(t)=\bar\omega_p b_{0p}^{}(t)+\sum_{i'}f_{i'p}^{}b_{C_{i'}}^{}(t),\label{eq0c}
\end{align}
\end{subequations}
supplemented with the initial condition, $b_N^{}(0)=1$ and $b_{C_i}^{}(0)=b_{0p}^{}(0)=0$. Treating these equations as in the previous section, we can perform summations in Eqs.~(\ref{eq0}a,b) analytically, by using the continuous limit, $\sum_i\to\int\rho_C^{}dE_{C_i}$ and $\sum_p\to\int\bar\rho d\bar\omega_p$, where $\rho_C^{}$ and $\bar\rho$ are the densities of the acceptor's states, $E_{C_i}$, and of the fictitious boson bath.

Equations~(\ref{eq0}) are treated in the same way as in the previous section, by assuming that the density of states and the spectral function, $|\tilde V|^2\bar\rho$, and $|f|^2\rho_C\bar\rho$, are the energy independent (Markovian limit). First, we perform the Laplace transform, $b(t)\to\tilde b(E)$, Eq.~(\ref{laplace}).  Then, resolving Eq.~(\ref{eq0c}) for the amplitude $\tilde b_{0p}(E)$ and substituting it in the previous equation, we rewrite Eqs.~(\ref{eq0}a,b) (after the inverse Laplace transform, Eq.~(\ref{invlap})) as,
\begin{subequations}
\label{eq1}
\begin{align}
&i\dot{b}_{N}^{}(t)=\Big(E_N^{}-i{\Gamma\over2}\Big){b}_{N}^{}(t),\label{eq1a}\\
&i\dot{b}_{C_i}^{}(t) =\Big(E_{C_i}^{}-i{\gamma_R^{}\over2}\Big)b_{C_i}^{}(t)+\tilde V b_{N}^{}(t),\label{eq1b}
\end{align}
\end{subequations}
where, $\Gamma=2\pi|\tilde V|^2\rho_C^{}$, is the rate of the charge separation and, $\gamma_R^{}=(2\pi)^2|f|^2\rho_C\bar\rho\Delta=1/\tau$, is the rate of the entire cycle (charge restoration). Here, $\Delta\simeq\Gamma$, is a width of the energy distribution of a  tunneling electron in the acceptor. Both rates are phenomenological parameters, which are determined experimentally ($1/\Gamma\sim ps$ and $1/\gamma_R\sim \mu s$).

Now we convert Eqs.~(\ref{eq0c}), (\ref{eq1}) into the Master equations for the density matrix of the system, defined as,
\begin{align}
&\sigma_{NN}^{}(t)=|b_{N}^{}(t)|^{2},~~
\sigma_{CC}^{}(t)=\sum_i|b_{C_i}^{}(t)|^{2},\nonumber\\
&\sigma_{00}^{}(t)=\sum_p|b_{0p}^{}(t)|^2.
\label{eq2}
\end{align}
Using the same procedure as in the previous section, we find,
\begin{subequations}
\label{eq3}
\begin{align}
&\dot\sigma_{NN}^{}(t)=-\Gamma \sigma_{NN}^{}(t),\label{eq3a}\\
&\dot\sigma_{CC}^{}(t) =\Gamma\sigma_{NN}^{}(t)-\gamma_R^{}\sigma_{CC}^{}(t),\label{eq3b}\\
&\dot\sigma_{00}^{}(t)=\gamma_R^{}\sigma_{CC}^{}(t).\label{eq3c}
\end{align}
\end{subequations}
Indeed, the first equation is obtained by taking imaginary part of Eq.~(\ref{eq1a}), multiplied by $b_N^*(t)$. The same procedure for Eq.~(\ref{eq1b}) yields,
\begin{align}
&i\dot\sigma_{CC}^{}(t) =-i\gamma_R^{}\sigma_{CC}^{}(t)+2i\tilde V\,{\rm Im}\big[b_N(t)\sum_ib_{C_i}^*(t)\big],
\label{eq4}
\end{align}
where $\tilde V\sum_ib_{C_i}^*(t)=i{\Gamma\over2}b_N^{}(t)$, as obtained
from Eqs.~(\ref{eq0a}) and (\ref{eq1a}). Substituting it into Eq.~(\ref{eq4}), we arrive to Eq.~(\ref{eq3b}). Last Eq.~(\ref{eq3c}) is obtained in a similar way.

Solving Eqs.~(\ref{eq3}) for the initial condition, $\sigma_{NN}^{}(0)=1$ and $\sigma_{CC}^{}(0)=\sigma_{00}^{}(0)=0$, we find
\begin{align}
&\sigma_{NN}^{}(t)=e^{-\Gamma t},~~\sigma_{CC}^{}(t)={e^{-\gamma_R^{} t}-e^{-\Gamma t}\over \Gamma-\gamma_R^{}}\Gamma\simeq e^{-\gamma_R^{}t},\nonumber\\
&\sigma_{00}^{}(t)=1-{\Gamma e^{-\gamma_R^{}t}-\gamma_R^{}e^{-\Gamma t}\over \Gamma-\gamma_R^{}}\simeq1-e^{-\gamma_R^{}t}.
\end{align}
As expected, the primary charge separation occurs during very small time-interval, $\sim\Gamma^{-1}$, whereas the charge restoration proceeds very slowly, $\sim\gamma_R^{-1}$.

\subsection{Complete exciton dynamics for one-site antenna.}

\begin{figure}[h]
\includegraphics[width=8cm]{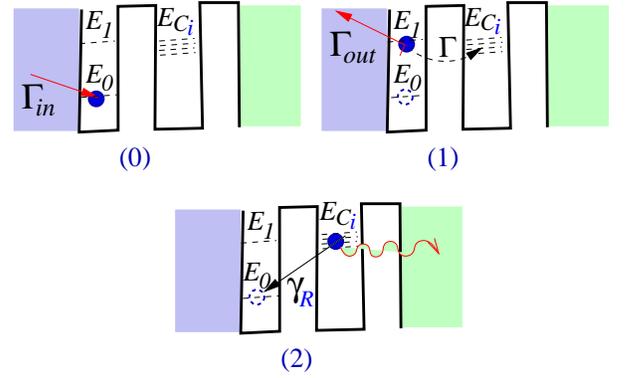}
\caption{Electron states for one-site antenna: (0) Exciton creation; (1) Photon emission and charge separation; (2) restoration of the donor's neutrality by emission of energy (fictitious boson) to the RC.}
\label{fig4}
\end{figure}

Now we can write the Master equations for an entire cycle, by combining the exciton creation with the charge separation and restoration dynamics. We start with an example of one-site antenna. ($N=1$), Fig.~\ref{fig4}.

The system is described by the Hamiltonian, ${\cal H}_1$,
\begin{align}
&{\cal H}_1=E_1\hat a_1^\dagger \hat a_1^{}+\sum_iE_{C_i}^{}\hat a_{C_i}^\dagger\hat a_{C_i}^{}+\sum_k\omega_k\hat C_k^\dagger \hat C_k^{}\nonumber\\
&+\sum_p\bar\omega_p^{}\hat F_p^\dagger \hat F_p^{}+\Big[\sum_k g_k\hat B_1^\dagger\hat C_k\nonumber\\
&+\sum_i\Big(\tilde V_i^{}\hat a_{C_i}^\dagger \hat a_N^{}
+\sum_p f_{ip}^{}\hat a_{0}^{\dagger}\hat a_{C_i}\hat F_p^{\dagger}\Big)+H.c.\Big],
\label{ap1}
\end{align}
which combines two Hamiltonians considered before, Eq.~(\ref{apb1}) and Eq.~(\ref{ham1}), for $N=1$.

Using the same procedure as in the previous sections, we convert the total wave function to particle-number-resolved Master equations for the reduced density-matrix, $\sigma_{jj}^{(\ell)}(t)$, where, $\ell$, denotes a number of fictitious bosons, emitted by time $t$ and representing a number of cycles (c.f. with Eqs.~(\ref{sigmas}) and (\ref{phot1}) for the reduced density matrix, which includes a number of emitted photons). Note that in our previous example, Eqs.~(\ref{eq2}) and (\ref{eq3}), the donor has been detached from incoming photons, so only one fictitious boson can be emitted ($\ell=1$).

The resulting Master equations have the following form,
\begin{subequations}
\label{apbb12}
\begin{align}
& \dot{\sigma}_{00}^{(\ell)}=-\Gamma_{in}\,\sigma_{00}^{(\ell)}
+\Gamma_{out}\sigma_{11}^{(\ell)}+\gamma_R^{}\sigma_{22}^{(\ell-1)}, \label{apbb12a} \\
&\dot{\sigma}_{11}^{(\ell)}=-(\Gamma_{out}+\Gamma)\,\sigma_{11}^{(\ell)}
+\Gamma_{in}\sigma_{00}^{(\ell)},
\label{apbb12b} \\
&\dot{\sigma}_{22}^{(\ell)}=-\gamma_R^{}\,\sigma_{22}^{(\ell)}
+\Gamma\sigma_{11}^{(\ell)},
\label{apbb12c} \\
\end{align}
\end{subequations}
where $\Gamma_{in,out}$ are given by Eq.~(\ref{gammas}). Respectively, the total probabilities, $\sigma_{jj}^{}(t)=\sum_{\ell}\sigma_{jj}^{(\ell)}(t)$, are given by the same Eqs.~(\ref{apbb12}), traced over $\ell$.

Equations~(\ref{apbb12}) have a form of classical rate equations, although they have been derived pure quantum mechanically. These equations have a clear interpretation in terms of the ``loss'' and ``gain'' processes (borrowing the terminology of the classical Boltzmann equation).  The off-diagonal matrix elements (coherences), are absent in Eqs.~(\ref{apbb12}), since all transitions take place between discrete and continuous (Markovian) states. However, in the case of transitions between different discrete states, off-diagonal density-matrix elements will occur in the equations of motion. This will be illustrated in following examples.

As we mentioned above, the emission of fictitious bosons, introduced phenomenologically for the restoration of the donor's neutrality, represents, in fact, the energy transfer to the RC. Therefore, Eqs.~(\ref{apbb12}) can be used to evaluate the corresponding energy current to the RC \cite{dubi}. Indeed, the latter is given by $I_{en}(t)=E_1\sum_\ell \dot P_\ell(t)$, where $P_\ell (t)=\sum_{j}\ell \sigma^{(\ell)}_{jj}(t)$ is the probability of finding $\ell$ emitted bosons by time $t$. Using Eqs.~(\ref{apbb12}), we easily obtain that,
\begin{align}
I_{en}(t)=E_1\,\gamma_R^{}\,\sigma_{22}^{}(t),
\label{cur2}
\end{align}
where $\sigma_{22}^{}(t)$ is the occupation probability of the acceptor. It is determined by tracing Eqs.~(\ref{apbb12}) over $\ell$,
\begin{subequations}
\label{apbb13}
\begin{align}
& \dot{\sigma}_{00}^{}=-\Gamma_{in}\,\sigma_{00}^{}
+\Gamma_{out}\sigma_{11}^{}+\gamma_R^{}\sigma_{22}^{}, \label{apbb13a} \\
&\dot{\sigma}_{11}^{}=-(\Gamma_{out}+\Gamma)\,\sigma_{11}^{}
+\Gamma_{in}\sigma_{00}^{},
\label{apbb13b} \\
&\dot{\sigma}_{22}^{}=-\gamma_R^{}\,\sigma_{22}^{}
+\Gamma\sigma_{11}^{}.
\label{apbb13c} \\
\end{align}
\end{subequations}

Consider a steady-state limit ($t\to\infty$), where $\dot\sigma(t)\to0$. In this case Eqs.~(\ref{apbb13}) become a system of algebraic equations for $\bar\sigma_{jj}=\sigma_{jj}(t\to\infty)$. Taking into account that $\bar\sigma_{00}=1-\bar\sigma_{11}-\bar\sigma_{22}$, we can easily resolve these equations, thus obtaining for the steady-state probabilities,
\begin{align}
&\bar\sigma_{00}={\gamma_R^{}\,(\Gamma+\Gamma_{out})\over \Gamma\Gamma_{in}+\gamma_R^{}\,(\Gamma+\Gamma_{in}+\Gamma_{out})},\nonumber\\
&\bar\sigma_{11}={\gamma_R^{}\Gamma_{in}\over \Gamma\Gamma_{in}+\gamma_R^{}(\Gamma+\Gamma_{in}+\Gamma_{out})},\nonumber\\
&\bar\sigma_{22}={\Gamma\,\Gamma_{in}\over \Gamma\Gamma_{in}+\gamma_R^{}(\Gamma+\Gamma_{in}+\Gamma_{out})},
\label{apbb14}
\end{align}
where, $\Gamma_{in,out}$, are given by Eq.~(\ref{gammas}).

In the case of $\gamma_R =0$, one finds that $\bar\sigma_{00}=\bar\sigma_{11}=0$, since there is no charge restoration. In the limit of $\bar n\to\infty$ (infinitely strong light intensity), one obtains that $\bar\sigma_{00}=\bar\sigma_{11}=\gamma_R^{}/\Gamma+{\cal O}(\gamma_R^{}/\Gamma)^2$. Respectively, $\bar\sigma_{22}=1-2\gamma_R^{}/\Gamma+{\cal O}(\gamma_R/\Gamma)^2$. This corresponds to equal occupation of ground and exited states of the donor,
$\bar\sigma_{00}=\bar\sigma_{11}=\gamma_R/\Gamma$.

\section{Exciton transport in $N$-site antenna}

\subsection{Master equations in general case.}

Now we extend our treatment on the $N$-site antenna chain, coupled with the electromagnetic field, describing photo-absorption and fluorescence, and with fictitious boson bath, describing the donor charge restoration,  Figs.~\ref{fig1} and \ref{fig2}.  The total Hamiltonian, describing this system can be written as,
\begin{align}
{\cal H}_N&=\sum_k\omega_k\hat C_k^\dagger \hat C_k^{}+\sum_{m=1}^{N}E_m\hat B_m^\dagger \hat B_m^{}+\sum_iE_{C_i}^{}\hat a_{C_i}^\dagger\hat a_{C_i}^{}
\nonumber\\
&~~~~~~~~~~~~~~~~~~~~~~~~~~~~~~
+\sum_p\bar\omega_p\hat F_p^\dagger \hat F_p^{}+H_{int},
\label{ham}
\end{align}
where $\hat B_m^\dagger =\hat a_m^\dagger \hat a_{0m}^{}$ is an exciton creation operator on the site $m$. Without loosing generality, we assume that the ground state energy of all sites $m=1,\ldots N$ equals zero. All notations are the same as in Eq.~(\ref{ap1}).
The interaction  term (in the rotated wave approximation) can be written as
\begin{align}
H_{int}&=\sum_{m=1}^N\sum_k g_k^{}\,\hat B_m^\dagger \hat C_k^{}
+\sum_{m=1}^{N-1} V_m^{}\,\hat B_{m+1}^\dagger \hat B_{m}^{}\nonumber\\
&+\sum_i\Big(\tilde V_i^{}\hat a_{C_i}^\dagger \hat a_N^{}
+\sum_p f_{ip}^{}\hat a_{0}^{\dagger}\hat a_{C_i}^{}\hat F_p^{\dagger}\Big)+H.c.
\label{hami}
\end{align}
Here the electromagnetic field is coupled with all sites of antenna. However, excitons can be generated only on the first antenna site, $m=1$, by photon absorption. All other sites, $m=2,\ldots N$, are coupled with empty photon reservoirs. Thus, the excitons occupying these sites can only decay by the fluorescence.

Note that the exciton commutation relation \cite{muc},
\begin{align}
[\hat B_m^{},\hat B_n^\dagger]=\delta_{mn}^{}(1-2\hat B_m^\dagger \hat B_n^{}),
\end{align}
guaranties that two or more excitons cannot occupy the same site. Therefore, the exciton motion along the antenna, describing by the Hamiltonian~(\ref{ham}), (\ref{hami}), is similar to that of the spinless electron transport trough the coupled quantum dots.

The corresponding Master equation can be derived from the time-dependent multi-particle Schr\"odinger equation, in the same way as discussed in previous examples. It represents a natural extension of Eqs.~(\ref{apbb12}) and (\ref{apbb13}), and can be written as (see Eq.~(75) of Ref.~[\onlinecite{gur}]):
\begin{align}
&\dot\sigma_{\alpha\alpha'}^{(\nu,\ell)} =  i\big({\cal E}_{\alpha'} - {\cal E}_{\alpha}\big) \sigma_{\alpha\alpha'}^{(\nu,\ell)}+i\sum_{\beta}\Big (\sigma_{\alpha\beta}^{(\nu,\ell)}
V_{\beta\to\alpha'}
-V_{\beta\to\alpha}\sigma_{\beta\alpha'}^{(\nu,\ell)}\Big)\nonumber\\
&-{1\over2}\sigma_{\alpha\alpha'}^{(\nu,\ell)}\sum_\beta
(\Gamma_{\alpha\to\beta}+\Gamma_{\alpha'\to\beta})
+\sum_{\beta,\beta'}\sigma_{\beta\beta'}^{(\nu',\ell')}
\Gamma_{\beta\to\alpha,\beta'\to\alpha'},
\label{d1}
\end{align}
where $|\alpha\rangle$, $|\beta\rangle$ enumerate all {\em discrete} multi-exciton states in the occupation number representation, and ${\cal E}_\alpha=\sum_{m\in\alpha}E_m$ is a total energy of the state, $|\alpha\rangle$.
The upper indices, $\nu$ and $\ell$, in the density matrix, $\sigma_{\alpha\alpha'}^{(\nu,\ell)}(t)$, denote the numbers of fluorescent photons and fictitious bosons emitted at time, $t$. Note, that in the last
(``gain'') term, $(\nu',\ell')=(\nu-1,\ell)$ or $(\nu',\ell')=(\nu,\ell-1)$, whenever emission of fluorescence photons or fictitious bosons takes place (c.f. with Eqs.~(\ref{apbb12})).

The second term in Eq.~(\ref{d1}) describes the direct exciton transitions between neighboring sites, $V_{\beta\alpha}= V_{m,m+1}\equiv V_m$, generated by dipole-dipole interaction (F\"orster mechanism). Obviously, this term does not appear in the rate equation~(\ref{apbb12}) for one-site antenna. One can realize that the first and second terms of Eq.~(\ref{d1}) represent the  commutator of the density-matrix with the Hamiltonian in the Lindbladt equation  \cite{lind}.

The remaining two terms represent loss and gain processes generated by: (a) coupling of the site ($1$) to photon bath with rates  $\Gamma_{\alpha,\beta}\equiv\Gamma_{in,out}$, Eqs.~(\ref{gammas}); (b) fluorescence of sites, $m=2,\ldots,N$, with the rate $\Gamma_{\alpha,\beta}\equiv\gamma$; (c) charge separation with subsequent emission of fictitious bosons, leading to restoration of the donor's neutrality with the rates $\Gamma_{\alpha,\beta}\equiv\Gamma,\gamma_R^{}$, respectively, Eqs.~(\ref{apbb12}). Note, that the gain processes, described by the fourth term in Eq.~(\ref{d1}), are generated by the {\em same} exciton transition leading to decay of the states $\beta,\beta'$ to the states $\alpha,\alpha'$.

By solving Eqs.~(\ref{d1}), we can determine probabilities of any multi-exciton occupations, as well as the fluorescent current (in energy units), $I_{fl}(t)$, and the current of energy, $I_{en}(t)$, transferred to the RC. Those are given by (c.f. with Ref.~[\onlinecite{gp,gur}]),
\begin{align}
&I_{fl}(t)=\sum_{\nu,\ell}
\sum_\alpha E_{m'}\nu\dot\sigma_{\alpha\alpha}^{(\nu,\ell)}(t)
=\gamma\sum_{\alpha'}E_{m'}\sigma_{\alpha'\alpha'}^{}(t),
\label{cur3}\\
&I_{en}(t)=E_N\sum_{\nu,\ell}
\sum_\alpha\ell\dot\sigma_{\alpha\alpha}^{(\nu,\ell)}(t)
=E_N^{}\gamma_R^{}\sum_{\alpha'}\sigma_{\alpha''\alpha''}^{}(t),
\label{cur3p}
\end{align}
where $\sigma_{\alpha\alpha}(t)=\sum_{\nu,\ell}\sigma_{\alpha\alpha}^{(\nu,\ell)}(t)$, are total probabilities, obtained from Eqs.~(\ref{d1}), and $m'=(2,\ldots ,N)\in \alpha$ denote the fluorescent sites (the first antenna site is excluded from our definition of fluorescent current). The indices,  $\alpha'$ and $\alpha''$, enumerate all multi-exciton states decaying by fluorescence or by emission of  fictitious bosons, respectively.

\subsection{Two-site antenna.}

\begin{figure}[h]
\includegraphics[width=8cm]{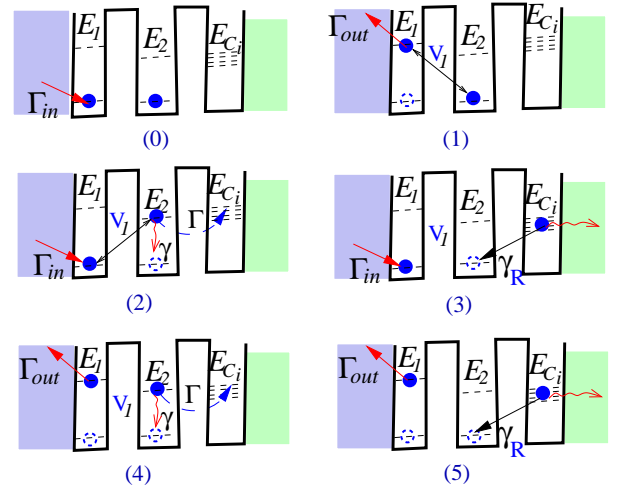}
\caption{Exciton states of the system. All allowed exciton transitions for each of the states are indicated.}
\label{fig5}
\end{figure}

As an example for application of Eq.~(\ref{d1}), we consider exciton transport through the two-site antenna, ($N=2$). First, we need to enumerate all possible exciton states of the system, $\{\alpha,\beta\}=\{0,1\ldots, 5\}$. These are shown in Fig.~\ref{fig5}. Note, that the exciton propagates coherently between the sites (1) and (2) due to the  dipole-dipole interaction, $V_1$. All other transitions are incoherent, where related transition rates are shown for each of the states.

Now we can rewrite explicitly the Master equations~(\ref{d1}) for the reduced density-matrix,
\begin{subequations}
\label{apb14}
\begin{align}
& \dot{\sigma}_{00}^{(\nu,\ell)}=-\Gamma_{in}\sigma_{00}^{(\nu,\ell)}
+\Gamma_{out}\sigma_{11}^{(\nu,\ell)}+\gamma\sigma_{22}^{(\nu-1,\ell)}\nonumber\\
&~~~~~~~~~~~~~~~~~~~~~~~~~~~~~~~~~~~~~~~~~~~~~~
+\gamma_R^{}\sigma_{33}^{(\nu,\ell-1)} \label{apb14a}, \\
&\dot{\sigma}_{11}^{(\nu,\ell)}=i V_1(\sigma_{12}^{(\nu,\ell)}-\sigma_{21}^{(\nu,\ell)})
-\Gamma_{out}\sigma_{11}^{(\nu,\ell)}
+\Gamma_{in}\sigma_{00}^{(\nu,\ell)}\nonumber\\
&~~~~~~~~~~~~~~~~~~~~~~~~~~~~~
+\gamma\,\sigma_{44}^{(\nu-1,\ell)}+\gamma_R^{}\sigma_{55}^{(\nu,\ell-1)},
\label{apb14b} \\
&\dot{\sigma}_{22}^{(\nu,\ell)}=i V_1(\sigma_{21}^{(\nu,\ell)}-\sigma_{12}^{(\nu,\ell)})
-(\gamma+\Gamma+\Gamma_{in})\sigma_{22}^{(\nu,\ell)}\nonumber\\
&~~~~~~~~~~~~~~~~~~~~~~~~~~~~~~~~~~~~~~~~~~~~~~
+\Gamma_{out}\sigma_{44}^{(\nu,\ell)},
\label{apb14c} \\
&\dot{\sigma}_{33}^{(\nu,\ell)}=-(\Gamma_{in}+\gamma_R^{})\sigma_{33}^{(\nu,\ell)}
+\Gamma_{out}\sigma_{55}^{(\nu,\ell)}+\Gamma\sigma_{22}^{(\nu,\ell)},
\label{apb14d}\\
&\dot{\sigma}_{44}^{(\nu,\ell)}=-(\gamma+\Gamma_{out}+\Gamma)\sigma_{44}^{(\nu,\ell)}
+\Gamma_{in}\sigma_{22}^{(\nu,\ell)}, \label{apb14e}\\
&\dot{\sigma}_{55}^{(\nu,\ell)}=-(\Gamma_{out}+\gamma_R^{})\sigma_{55}^{(\nu,\ell)}
+\Gamma_{in}\sigma_{33}^{(\nu,\ell)}+\Gamma\sigma_{44}^{(\nu,\ell)},\label{apb14f}\\
&\dot{\sigma}_{12}^{(\nu,\ell)}=i(E_{2}-E_1)\sigma_{12}^{(\nu,\ell)}
+iV_1(\sigma_{11}^{(\nu,\ell)}-\sigma_{22}^{(\nu,\ell)})
\nonumber\\
&~~~~~~~~~~~~~~~~~~~~~~~~~~~
-{\Gamma+\gamma+\Gamma_{in}+\Gamma_{out}\over2}\sigma_{12}^{(\nu,\ell)}.
\label{apb14g}
\end{align}
\end{subequations}
The total probabilities, $\sigma_{\alpha\alpha'}(t)=\sum_{\nu,\ell}\sigma_{\alpha\alpha'}^{(\nu,\ell)}(t)$,
are given by the same Eqs.~(\ref{apb14}) by removing the upper indices.
One can easily verify that these equations display the probability conservation, \begin{align}
\sum_{\alpha=0}^5\sigma_{\alpha\alpha}(t)=1.
\label{repl00}
\end{align}

Solving Eqs.~(\ref{apb14}), we can find the probabilities for one and two excitons inside the antenna,
\begin{align}
&P_1(t)=\sigma_{11}(t)+\sigma_{22}(t)+\sigma_{55}(t),\nonumber\\
&P_2(t)=\sigma_{44}(t).
\label{ocprob}
\end{align}
Using Eqs.~(\ref{cur3}) and (\ref{cur3p}), we find that
the fluorescent current (on the second site, Fig.~\ref{fig5}) is given by,
\begin{align}
I_{fl}(t)
=\gamma E_2[\sigma_{22}(t)+\sigma_{44}(t)],
\label{cur44}
\end{align}
and the energy current to the RC is,
\begin{align}
I_{en}(t)
=\gamma_R^{} E_2[\sigma_{33}(t)+\sigma_{55}(t)].
\label{cur33}
\end{align}

Consider now the steady-state limit, $t\to\infty$. Since in this limit  $\dot\sigma_{\alpha\alpha'}\to 0$, Eqs.~(\ref{apb14}) become a system of algebraic equations for $\bar\sigma\equiv \sigma (t\to\infty )$.
Using the probability conservation, Eq.~(\ref{repl00}), we can rewrite these equations as,
\begin{subequations}
\label{st14}
\begin{align}
&\bsigma_{00}^{}+\bsigma_{11}^{}+\bsigma_{22}^{}+\bsigma_{33}^{}+\bsigma_{44}^{}
+\bsigma_{55}^{}=1,\label{st14a} \\
&i V_1(\bsigma_{12}^{}-\bsigma_{21}^{})
-\Gamma_{out}\bsigma_{11}^{}
+\Gamma_{in}\bsigma_{00}^{}\nonumber\\
&~~~~~~~~~~~~~~~~~~~~~~~~~~~~~
+\gamma\,\bsigma_{44}^{}+\gamma_R^{}\bsigma_{55}^{}=0,
\label{st14b} \\
&i V_1(\bsigma_{21}^{}-\bsigma_{12}^{})
-(\gamma+\Gamma+\Gamma_{in})\bsigma_{22}^{}
+\Gamma_{out}\bsigma_{44}^{}=0,
\label{st14c} \\
&-(\Gamma_{in}+\gamma_R^{})\bsigma_{33}^{}
+\Gamma_{out}\bsigma_{55}^{}+\Gamma\bsigma_{22}^{}=0,
\label{st14d}\\
&\-(\gamma+\Gamma_{out}+\Gamma)\bsigma_{44}^{}
+\Gamma_{in}\bsigma_{22}^{}=0, \label{st14e}\\
&-(\Gamma_{out}+\gamma_R^{})\bsigma_{55}^{}
+\Gamma_{in}\bsigma_{33}^{}+\Gamma\bsigma_{44}^{}=0,\label{apb14f}\\
&i(E_{2}-E_1)\bsigma_{12}^{}
+iV_1(\bsigma_{11}^{}-\bsigma_{22}^{})
\nonumber\\
&~~~~~~~~~~~~~~~~~~~~~~~~~
-{\Gamma+\gamma+\Gamma_{in}+\Gamma_{out}\over2}\bsigma_{12}^{}=0.
\label{st14g}
\end{align}
\end{subequations}

\begin{figure}[h]
\includegraphics[width=8cm]{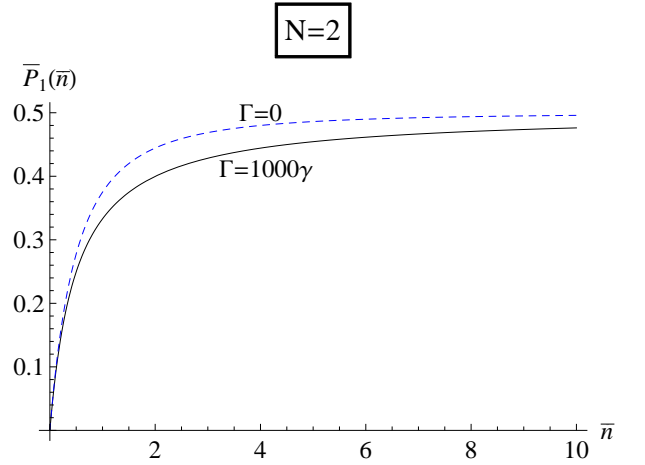}
\caption{Probability of trapping one exciton, $\bar P_1(\bar n)$ as a function of light intensity ($\bar n$ is a number of photons absorbed by the first site for 1 fs). Solid line (black) corresponds to $\Gamma=10^3\gamma$=1/ps, whereas dashed line (blue) corresponds to $\Gamma=0$,  corresponding to no-coupling of antenna with RC. Other parameters are:
$\gamma$=1/ns, $\gamma_R=10^{-3}\gamma=1/\mu$s, $V_1=10^3\gamma$=1/ps.}
\label{fig5p}
\end{figure}

\begin{figure}[h]
\includegraphics[width=8cm]{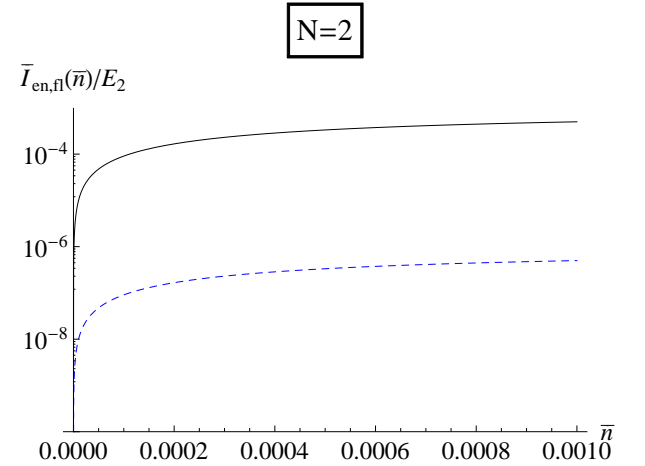}
\caption{Steady-state energy current, $\bar I_{en}$ (solid, black) and fluorescent current $\bar I_{fl}$ (dashed, blue) in units of the donor energy $E_2$ per ns, as a function of light intensity, $\bar n$, for $\gamma_R=10^{-3}\gamma=1 \mu$s, $V_1=\Gamma=10^3\gamma$=1 ps.}
\label{fig5pp}
\end{figure}

Steady-state probability for single exciton occupation, $\bar P_1(\bar n)$, and steady-state currents, $\bar I_{fl,en}(\bar n)$, (Eqs.~(\ref{ocprob}), (\ref{cur44}) and (\ref{cur33}) for $t\to\infty$), are shown in Figs.~\ref{fig5p} and \ref{fig5pp} as functions of $\bar n$ (light intensity). Note, that $\bar n\gamma=\Gamma_{in}$, Eq.~(\ref{gammas}), is a number of photons absorbed by the first site per unit time. Here and in what follows, we choose for illustrative examples some generic values of parameters, not necessary related to a specific system, namely, $\gamma$=1/ns (fluorescent rate),  $\gamma_R=1/\tau=10^{-3}\gamma=1/\mu$s (charge restoration rate), $V_1=10^3\gamma$=1/ ps (dipole-dipole coupling between sites), and $\Gamma=10^3\gamma$=1/ps (charge separation rate).

Figure~\ref{fig5p}, displays the probability, $\bar P_1(\bar n)$, of trapping one exciton for two values of
$\Gamma=10^3\gamma$=1/ps (solid line) and $\Gamma=0$ (dashed line). The latter  corresponds to no-coupling of antenna with the RC.
One finds from this figure that the single-exciton occupation is saturated with increase of $\bar n$. As expected, the saturation takes place at equal occupation of ground and excited states for $\bar n\to\infty$, Eq.~(\ref{st}). The probability of double-exciton occupation, $P_2(\bar n)$, Eq.~(\ref{ocprob}) is not shown here, since it is very small for $N=2$. However, already for $N=3$, this quantity, $P_2(\bar n)$, becomes quite large, as demonstrated below.

The steady-state energy current to the RC, $\bar I_{en}$ (solid line, black) and  the fluorescent current of the second site, $\bar I_{fl}$ (dashed line, blue) in units of donor energy ($E_2$) per $1/\gamma$=1 ps, are shown in Fig.~\ref{fig5pp}, as a function of $\bar n$.
One finds from this figure that the currents reach saturation already for a very small $\bar n$, where less than one photon is absorbed at the first site during 1 $\mu$s. This rate is of the order of optimal photo-absorption rate, when no more than one exciton can be accumulated inside the antenna during one cycle. (The  time of cycle is $\tau=1/\gamma_R=\mu$s.) The origin of the saturation can be easily understood. Indeed, an increase of the light intensity results in generation of  additional excitons, trapped inside the antenna. These excitons will be eventually removed by fluorescence, so the energy current will not increase anymore.

\begin{figure}[h]
\includegraphics[width=8cm]{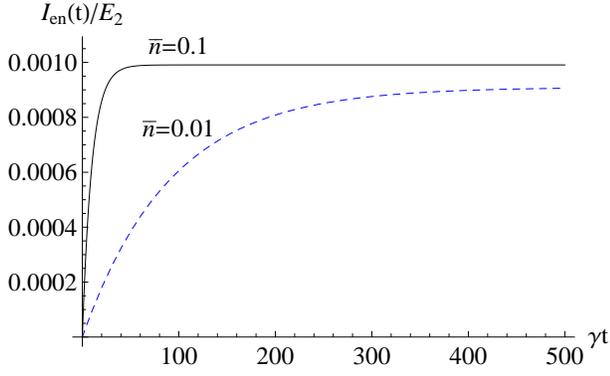}
\caption{Time-dependent energy current transferred to the RC for $\bar n=0.1$ (solid line, black) and $\bar n=0.01$ (dashed line, blue) in units of the donor energy $E_2$ per ns. Time, $t$, is taken in units of $1/\gamma$=1 ns. All parameters are the same as in Fig.~\ref{fig5pp}.}
\label{fig6pp}
\end{figure}

The time-dependent energy current, $I_{en}(t)$, Eq.~(\ref{cur33}), is displayed in Fig.~\ref{fig6pp} for $\bar n=0.1$ (solid line, black) and $\bar n=0.01$ (dashed line, blue). One finds that time for approaching the asymptotic limit increases when the intensity of light, $\bar n$, decreases. For an optimal intensity, fitted to a cycle period, the saturation time can be quite long.

Note that the fluorescent current, $\bar I_{fl}$, Eq.~(\ref{cur44}), is very small, in comparison with the energy current, $\bar I_{en}$, Fig.~(\ref{fig5pp}). On first sight, one could expect an opposite result. Indeed, $\bar I_{en}$ is proportional to $\gamma_R$, whereas $\bar I_{fl}$ is proportional to $\gamma$ (Eqs.~(\ref{cur33}), (\ref{cur44})), which greatly exceeds $\gamma_R$. However, the fluorescent current in the case of $N=2$ can take place only when the second site (donor) is occupied, and therefore it is very small. This, however, is not the case for $N>2$, as will be demonstrate below.

\subsection{Three-site antenna and exciton accumulation.}

\begin{figure}[h]
\includegraphics[width=8cm]{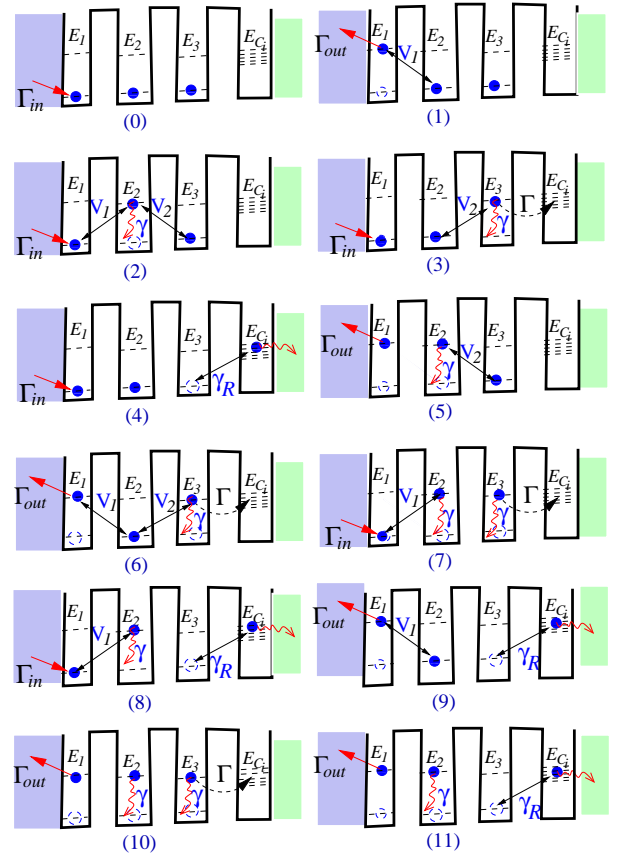}
\caption{Exciton states of the system. All allowed exciton transitions for each of the states are indicated.}
\label{fig6}
\end{figure}

Consider now a three-site antenna ($N=3$). All possible exciton states are shown in Fig.~\ref{fig6}. The state (0) displays the ground state (no excitons).
The states (1),(2),(3) display one-exciton states, (5),(6),(7) display two-exciton states, and (10) displays three-exciton stares. The states (4),(8),(9),(11) are the charge states when the donor is blocked.

The rate equations, describing the exciton dynamics, are obtained directly from Eq.~(\ref{d1}). We find,
\begin{subequations}
\label{apb15}
\begin{align}
& \dot{\sigma}_{00}=-\Gamma_{in}\sigma_{00}
+\Gamma_{out}\sigma_{11}+\gamma(\sigma_{22}+\sigma_{33})+\gamma_R^{}\sigma_{44}, \label{apb15a} \\
&\dot{\sigma}_{11}=i V_1(\sigma_{12}-\sigma_{21})-\Gamma_{out}\sigma_{11}
+\Gamma_{in}\sigma_{00}\nonumber\\
&~~~~~~~~~~~~~~~~~~~~~~~~~~~~
+\gamma(\sigma_{55}+\sigma_{66})+\gamma_R^{}\sigma_{99},
\label{apb15b} \\
&\dot{\sigma}_{22}=i V_1(\sigma_{21}-\sigma_{12})+i V_2(\sigma_{23}-\sigma_{32})\nonumber\\
&~~~~~~~~~
-(\Gamma_{in}+\gamma)\sigma_{22}
+\Gamma_{out}\sigma_{55}+\gamma\sigma_{77}+\gamma_R^{}\sigma_{88},
\label{apb15c} \\
&\dot{\sigma}_{33}=i V_2(\sigma_{32}-\sigma_{23})
-(\Gamma_{in}+\gamma+\Gamma)\sigma_{33}\nonumber\\
&~~~~~~~~~~~~~~~~~~~~~~~~~~~~~~~~~~~~~~~~
+\Gamma_{out}\sigma_{66}+\gamma\sigma_{77},
\label{apb15d}\\
&\dot{\sigma}_{44}=-(\Gamma_{in}+\gamma_R^{})\sigma_{44}
+\Gamma\sigma_{33}+\gamma\sigma_{88} +\Gamma_{out}\sigma_{99},
\label{apb15e}\\
&\dot{\sigma}_{55}=i V_2(\sigma_{56}-\sigma_{65})
-(\Gamma_{out}+\gamma)\sigma_{55}\nonumber\\
&~~~~~~~~~~~~~~~~~~~~~~~
+\Gamma_{in}\sigma_{22}+\gamma\sigma_{10,10}
+\gamma_R^{}\sigma_{11,11}\label{apb15f},\\
&\dot{\sigma}_{66}=iV_1(\sigma_{67}-\sigma_{76})
+iV_2(\sigma_{65}-\sigma_{56})\nonumber\\
&~~~~~~~~~~~~
-(\Gamma_{out}+\gamma+\Gamma)\sigma_{66}
+\Gamma_{in}\sigma_{33}+\gamma\sigma_{10,10},\label{apb15g}\\
&\dot{\sigma}_{77}=i V_1(\sigma_{76}-\sigma_{67})
-(\Gamma_{in}+2\gamma+\Gamma)\sigma_{77}\nonumber\\
&~~~~~~~~~~~~~~~~~~~~~~~~~~~~~~~~~~~~~~~~~~~~~
+\Gamma_{out}\sigma_{10,10},\label{apb15h}\\
&\dot{\sigma}_{88}=i V_1(\sigma_{89}-\sigma_{98})
-(\Gamma_{in}+\gamma+\gamma_R^{})\sigma_{88}\nonumber\\
&~~~~~~~~~~~~~~~~~~~~~~~~~~~~~~~~~~~~~
+\Gamma\sigma_{77}+\Gamma_{out}\sigma_{11,11},\label{apb15aa}\\
&\dot{\sigma}_{99}=i V_1(\sigma_{98}-\sigma_{89})
-(\Gamma_{out}+\gamma_R^{})\sigma_{99}+\Gamma_{in}\sigma_{44}
\nonumber\\
&~~~~~~~~~~~~~~~~~~~~~~~~~~~~~~~~~~~~~~~
+\Gamma\sigma_{66}+\gamma\sigma_{11,11},\label{apb15bb}\\
&\dot{\sigma}_{10,10}=-(\Gamma_{out}+2\gamma
+\Gamma)\sigma_{10,10}+\Gamma_{in}\sigma_{77},\label{apb15cc}\\
&\dot{\sigma}_{11,11}=-(\Gamma_{out}+\gamma+\gamma_R^{})\sigma_{11,11}
+\Gamma_{in}\sigma_{88}\nonumber\\
&~~~~~~~~~~~~~~~~~~~~~~~~~~~~~~~~~~~~~~~~~~~~~~~
+\Gamma\sigma_{10,10},\label{apb15dd}\\
&\dot{\sigma}_{12}=i(E_2-E_1)\sigma_{12}+iV_1(\sigma_{11}-\sigma_{22})
+i V_2\sigma_{13}
\nonumber\\
&~~~~~~~~~~
-{\Gamma_{in}+\Gamma_{out}+\gamma\over2}\sigma_{12}
+\gamma\sigma_{67}+\gamma_R^{}\sigma_{98},
\label{apb15i}\\
&\dot{\sigma}_{13}=i(E_3-E_1)\sigma_{13}+iV_2\sigma_{12}-i V_1\sigma_{23}\nonumber\\
&~~~~~~~~~~~~~~~~~~~~~~~~~~~
-{\Gamma_{in}+\Gamma_{out}+\Gamma+\gamma\over2}\sigma_{13},
\label{apb15j}\\
&\dot{\sigma}_{23}=i(E_3-E_2)\sigma_{23}
+iV_2(\sigma_{22}-\sigma_{33})-i V_1\sigma_{13}\nonumber\\
&~~~~~~~~~~~~~~~~~~~~~~
-{2\Gamma_{in}+2\gamma+\Gamma\over2}\sigma_{23}+\Gamma_{out}\sigma_{56},
\label{apb15k}\\
&\dot{\sigma}_{56}=i(E_3-E_2)\sigma_{56}
+iV_2(\sigma_{55}-\sigma_{66})+iV_1\sigma_{57}
\nonumber\\
&~~~~~~~~~~~~~~~~~~~~
-{2\Gamma_{out}+2\gamma+\Gamma\over2}\sigma_{56}
+\Gamma_{in}\sigma_{23},
\label{apb15l}\\
&\dot{\sigma}_{57}=i(E_3-E_1)\sigma_{57}
+iV_1\sigma_{56}-iV_2\sigma_{67}\nonumber\\
&~~~~~~~~~~~~~~~~~~~~~~~~~~~~
-{\Gamma_{in}+\Gamma_{out}+3\gamma+\Gamma\over2}\sigma_{57},
\label{apb15m}\\
&\dot{\sigma}_{67}=i(E_2-E_1)\sigma_{67}
+iV_1(\sigma_{66}-\sigma_{77})
-iV_2\sigma_{57}\nonumber\\
&~~~~~~~~~~~~~~~~~~~~~~~~~~
-{\Gamma_{in}+\Gamma_{out}+3\gamma+2\Gamma\over2}\sigma_{67},
\label{apb15n}\\
&\dot{\sigma}_{89}=i(E_1-E_2)\sigma_{89}
+iV_1(\sigma_{88}-\sigma_{99})\nonumber\\
&~~~~~~~~~~~~~~~~~
-{\Gamma_{in}+\Gamma_{out}+\gamma+2\gamma_R^{}\over2}\sigma_{89}+\Gamma\sigma_{76}.
\label{apb15o}
\end{align}
\end{subequations}
Here we omitted for brevity the sub-indices $\ell$, and $\nu$, indicating a number of fictitious bosons and fluorescence photons emitted (c.f. Eqs.~(\ref{apb14})). Solving Eqs.~(\ref{apb15}), we obtain the reduced density matrix of the system, $\sigma_{\alpha\alpha'}^{}(t)=\sum_{\nu,\ell}\sigma_{\alpha\alpha'}^{(\nu,\ell)}(t)$. Then, we can evaluate probabilities for one, two and three exciton configurations inside the antenna, Fig.~\ref{fig6},
\begin{align}
&P_1(t)=\sigma_{11}(t)+\sigma_{22}(t)+\sigma_{33}(t)+\sigma_{88}(t)
+\sigma_{99}(t),\nonumber\\
&P_2(t)=\sigma_{55}(t)+\sigma_{66}(t)+\sigma_{77}(t)+\sigma_{11,11}(t)
+\sigma_{99}(t),\nonumber\\
&P_3(t)=\sigma_{10,10}(t).
\end{align}

\begin{figure}[h]
\includegraphics[width=8cm]{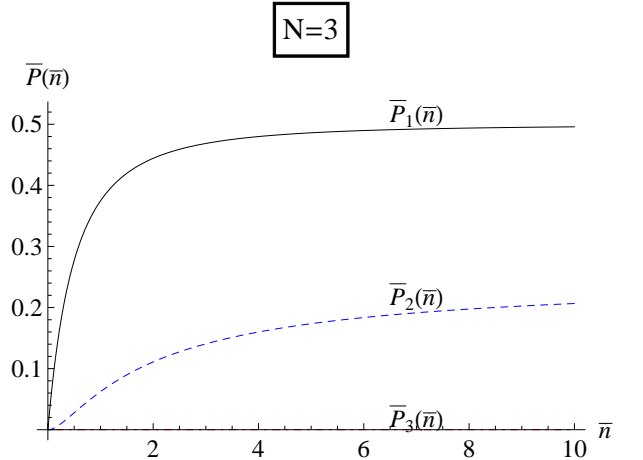}
\caption{Probability of trapping  one, two and three excitons inside the 3-site antenna at steady-state, $\bar P_{1}(\bar n)$ (solid, black), $\bar P_{2}(\bar n)$ (dashed, blue) and $\bar P_3(\bar n)$ (dot-dashed, red) as a function of light intensity, $\bar n$, for $\gamma_R=10^{-3}\gamma$, $V_1=V_2=10^3\gamma$, $\Gamma=10^3\gamma$, where $E_1=E_2=E_3$.}
\label{fig7}
\end{figure}

The results for steady-state, $\bar P=P(t\to\infty)$, are displayed in Fig.~\ref{fig7}.
In contrast with the two-site antenna, Fig.~\ref{fig5p}, the probability for two-exciton trapping is comparable with that for a one-exciton, $\bar P_1(\bar n)$. Three exciton trapping, $\bar P_3(\bar n)$, however, is negligibly small.

\begin{figure}[h]
\includegraphics[width=8cm]{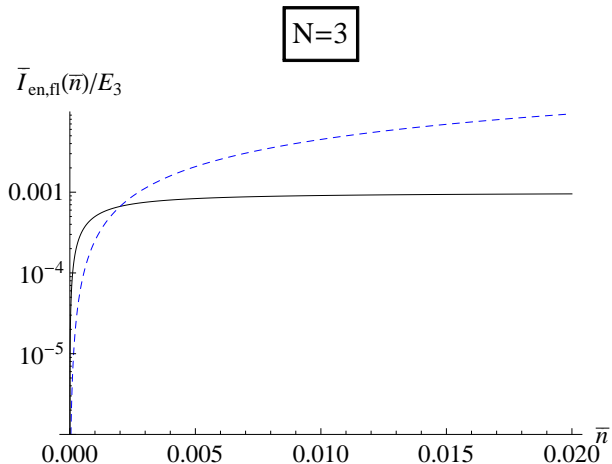}
\caption{Steady-state energy current $\bar I_{en}$ (solid, black) and fluorescent current $\bar I_{fl}$ (dashed, blue), in units of donor energy ($E_3$) per 1 ns, as a function of the light intensity, $\bar n$, where
for $\gamma_R=10^{-3}\gamma=1 \mu$s, $V_1=V_2=\Gamma=10^3\gamma$=1 ps.}
\label{fig8}
\end{figure}

The fluorescent current, $I_{fl}(t)$, Eqs.~(\ref{cur3}), from the second and the third sites, can be written explicitly as,
\begin{align}
I_{fl}(t)
&=\gamma[E_2\sigma_{22}(t)+E_3\sigma_{33}(t)+E_2\sigma_{55}(t)\nonumber\\
&+(E_2+E_3)\sigma_{77}(t)+E_2\sigma_{88}(t)\nonumber\\
&+(E_2+E_3)\sigma_{10,10}(t)+E_2\sigma_{11,11}(t)].
\label{cur55}
\end{align}
Respectively, the energy current, transferred to the RC,  $I_{en}(t)$, (\ref{cur3p}), reads,
\begin{align}
I_{en}(t)
=\gamma_R^{} E_3[\sigma_{44}(t)+\sigma_{88}(t)+\sigma_{99}(t)+\sigma_{11,11}(t)].
\label{cur55p}
\end{align}
Both currents at steady-state, $\bar I=I(t\to\infty )$ (in units of donor energy $E_3$ per $1/\gamma$=1 ns), are displayed in Fig.~\ref{fig8} for aligned levels, $E_1=E_2=E_3$, and for the same parameters as in Fig.~\ref{fig7}. The energy current,  $\bar I_{en}$, is displayed by solid (black) line, and the fluorescence current, $\bar I_{fl}$, is shown by dashed (blue) line. One finds  that the currents are saturated for large $\bar n$. Here, in contrast with the case $N=2$, Fig.~\ref{fig5pp}, the fluorescent current strongly dominates for large $\bar n$.

However, for small $\bar n$, the energy  current dominates over the fluorescent current. Note, that in that region both currents, $\bar I_{en,fl}(\bar n)$, are very sensitive to $\bar n$. It implies that a reliable analysis of the optimal regime for the LHC performance cannot be done without full account of all relevant physical effects. Therefore, our microscopic approach, which provides such an account, can be very relevant to determine conditions when the LHC works with a maximal efficiency.

\section{Discussion}

Although most investigations of energy (exciton) transport in the LHCs concentrate on one-exciton motion along the antenna, we demonstrated that it is not sufficient for a consistent description of exciton dynamics, in particular for an account of the NPQ-type processes. Therefore, we extend the Hamiltonian by including additional parts, describing the exciton creation and the fluorescence, through the interaction with the electromagnetic field, charge separation on the donor site, and the charge restoration after completing the corresponding cycle of the chemical reactions in the RC. The latter part is described phenomenologically, as an electron relaxation from the RC to the donor's ground state by emission of fictitious bosons. This process represents the energy transfer to the RC.

We have to emphasize that our effective description of the charge restoration involves the relaxation of the same electron, coming from the  charge separation. In reality, the process of reduction of the RC donor is more complicated. Nevertheless, this issue is not a deficiency of our model, since different electrons are indistinguishable and its origin is not relevant for the description of LHC dynamics. The relevant quantity is only the duration of the cycle, which is determined experimentally. An important assumption in our treatment, however, is the exponential relaxation process, which models  reduction (charge restoration) of the RC donor. In principle, the relaxation  could be of a different type, like a power-low. Then, it would imply a different Hamiltonian term describing the fictitious bosons.

We consider our approach as a general framework for constructing closed Master equations for complete description of the LHC. Additional effects can be accounted for in the same way as explained above, by adding the corresponding terms in the Master equations. For instance, any site of the antenna can be coupled with a sink, modeling the NPQ processes, in addition to the fluorescence.

In a similar way, one can introduce vibrational modes of the antenna sites in the Hamiltonian, together with environmental noise. These effects can play a very important role in the energy transport \cite{brumer,plenio}. Indeed, the antenna levels in general, are non-aligned. If the energy difference between sites is larger than the inter-site coupling, then transitions will be strongly reduced (Anderson localization). In this case, the quasi-resonant vibrational modes can close the gap. Perfect matching is not necessary if the environmental noise is taken into account \cite{plenio}.

The inclusion of vibrational modes together with the noise in our Master equation is not considered in the present paper. This issue will be discussed in a separate work, by treating these effects in a framework of a dichotomic (telegraph) noise. Such a procedure would double the number of rate equations, without additional complications \cite{shapiro,bg,ga}.

In general, our cycled multi-exciton Mater equations represent a more detailed description of the LHC dynamics. Therefore, by using these equations one can evaluate important effects, which cannot be treated by other methods. However, the number of multi-exciton states strongly increases with a number of antenna sites, as we have already seen by comparing $N=2$ with $N=3$ cases. This is a general problem for any treatments involving the density-matrix. There exists, however, an alternative, single-electron approach for mesoscopic transport \cite{single}, which can treat the wave function, instead of the density-matrix, even in the presence of noise \cite{ga}. Then, the number of equations will be drastically reduced.  An extension of this approach to the LHC is a topic of a future investigation.

\section*{Acknowledgment}
This work was carried out under the auspices of the National Nuclear Security
Administration of the U.S. Department of Energy at Los Alamos National Laboratory
under Contract No. DE-AC52-06NA25396.  R.T.S. acknowledges support from the LDRD program at LANL.

\end{document}